\documentclass[aps,pre,reprint,superscriptaddress]{revtex4-1}

\pdfoutput=1

\usepackage{graphicx}
\usepackage{amssymb}
\usepackage{amsmath}
\usepackage{xcolor}
\usepackage{pdfpages}
\usepackage{pgffor}
\usepackage{CJK}
\usepackage{multirow}

\makeatletter
\AtBeginDocument{\let\LS@rot\@undefined}
\makeatother

\begin{document}

\title{Atomic Origins of Friction Reduction in Metal Alloys} 

\author{Shengfeng Cheng}
\email{chengsf@vt.edu}
\affiliation{Department of Physics, Center for Soft Matter and Biological Physics, Macromolecules Innovation Institute, and Department of Mechanical Engineering, Virginia Polytechnic Institute and State University, Blacksburg, VA 24061, USA}
\author{Michael Chandross}
\email{mechand@sandia.gov}
\affiliation{Material, Physical and Chemical Sciences Center, Sandia National Laboratories, Albuquerque, NM 87185, USA }

\date{\today}

\begin{abstract}
We present the results of large scale molecular dynamics simulations aimed at understanding the origins of high friction coefficients in pure metals, and their concomitant reduction in alloys and composites. We utilize a series of targeted simulations to demonstrate that different slip mechanisms are active in the two systems, leading to differing frictional behavior. Specifically, we show that in pure metals, sliding occurs along the crystallographic slip planes, whereas in alloys shear is accommodated by grain boundaries. In pure metals, there is significant grain growth induced by the applied shear stress and the slip planes are commensurate contacts with high friction. However, the presence of dissimilar atoms in alloys suppresses grain growth and stabilizes grain boundaries, leading to low friction via grain boundary sliding.
\end{abstract}


\maketitle 

\section{Introduction}

A deep understanding of the tribology of gold contacts has been desired for hundreds of years, dating back to at least the 1700s, when there were documented investigations regarding the wear of gold coins and its potentially devastating effect on the economy of the United Kingdom.\cite{chanston74} Gold contacts have applications ranging from stereo connectors to sliding electrical contacts in wind turbines because of the high conductivity, ductility, and chemical inertness of gold. High adhesion and friction, however, can limit its applicability. While alloying can reduce friction,\cite{Haseeb2008,Su2013Wear} the generally accepted explanation relating this to higher ductility in a pure metal versus increased hardness in an alloy is based on empirical evidence only. An understanding of the mechanisms of friction reduction in alloys at the atomic scale is valuable for the design of materials with multiple potential applications in modern electronics and nanotechnology.\cite{Han2008}

Extensive research over the past several decades has provided insight into the mechanical properties of polycrystalline metals and alloys.\cite{Kumar2003,Wolf2005,Swygenhoven2006,Meyers2006,Spearot2019} It is now understood that in coarse-grained metals plastic deformation is dominated by dislocation-related processes. As grain sizes are reduced, the transmission of dislocations across grains is suppressed by interactions with grain boundaries and an increasing density of defects, including vacancies, inclusions, dislocations, and stacking faults. This leads to the ubiquitous Hall-Petch effect where the yield strength of a polycrystalline metal increases as grains get smaller.\cite{Swygenhoven2002Science,Cordero2016} However, when the grain sizes of a metal are reduced to the order of tens of nanometers, the density of grain boundaries becomes large and intergranular motion (including diffusion, sliding, and rotation) starts to control deformation. This results in softening of the metal, often referred to as the inverse Hall-Petch behavior or Hall-Petch softening.\cite{Schiotz1998,Cordero2016} As a result, there exists an intermediate grain size at which a metal exhibits maximum strength.\cite{Yip1998,Schiotz2003,Schiotz2004,Chandross2020,Gupta2020}

The transition of the dominant deformation mechanisms from dislocation to grain boundary-mediated plasticity with decreasing grain sizes in polycrystalline metals also plays a role in their frictional behavior. While reports indicate that coefficients of friction generally become lower as crystal grains are made smaller,\cite{Farhat1996,Jeong2001,Ful2012TL,Ma2013TL,Misochenko2017TL} the applied load,\cite{Ful2012TL} sliding speed,\cite{Shafiei2008} and lubrication condition \cite{Moshkovich2011} in a specific system may complicate the correlation between friction and grain size. There is also a report showing that nanocrystalline and coarse-grained copper have comparable coefficients of friction.\cite{Li2012Wear} On the other hand, many reports show that friction can be reduced by treating a metal to induce a nanocrystalline surface layer\cite{Zhang2006Wear,Wen2012,Amanov2012,Wang2013MSEA,She2015} with the explanation being that a surface layer with nanosized grains has higher hardness because of the Hall-Petch effect, and this higher hardness reduces wear and friction. More recent work shows that gradients in grain size at a metal surface lead to a similar effect.\cite{Chen2016SciAdv,Chen2018ACSAMI,Chen2020Friction,Chen2020SM}

The evolution of the microstructure at a metal surface under shear stress controls its frictional response, and this has recently been the subject of a great deal of research.\cite{Dienwiebel2007TL,Kato2010,Prasad2011,Padilla2013,Prasad2020,Argibay2016,Argibay2017,Cao2017TI,Greiner2018,Greiner2019,Eder2020} Normal and shear stresses in the contact zone affect the size of surface grains, which in turn influences the friction and stress distribution in this region.\cite{Argibay2017} This enables the development of a framework to predict the macroscale friction response (e.g., a low-to-high friction transition) of a metal surface, depending on material properties and loading conditions.\cite{Argibay2017} Controlling the microstructural evolution at a sliding interface therefore seems to be an important route to controlling wear \cite{Curry2018} and friction.\cite{Chandross2018,Jones2020} Specifically, this means stabilizing grain boundaries in the surface layer to maintain a desired nanocrystalline microstructure, as can occur with alloying \cite{Curry2018,Jones2020} or cooling the contact to cryogenic temperatures.\cite{Chandross2018}

The correlation between microstructural evolution at a sliding interface and its friction is then the key to understanding the mechanisms of friction reduction in alloys.\cite{Haseeb2008,Su2013Wear} While grains in alloys can be more stable when grain growth is mitigated\cite{Devaraj2019}, in a pure metal, stress-induced grain growth is expected at the sliding contact. As a result, alloys and pure metals can exhibit different deformation mechanisms at the sliding interface, leading to different frictional responses.\cite{Argibay2017} This is similar to the change of friction in a pure polycrystalline metal as the grain size is varied, leading to a transition in the dominant mode of plastic deformation.\cite{Scherge2003,Shakhvorostov2006,Shakhvorostov2007,Prasad2011,Padilla2013,Prasad2020,Hinkle2020,Gupta2020} Here, we use molecular dynamics (MD) simulations to examine these similarities by comparing the frictional response of ultra-nanocrystalline samples of pure Ag and an Ag-Cu alloy, in both tip-on-slab and slab-on-slab geometries. Clear evidence shows that metals and alloys indeed show different mechanisms of plastic deformation at the sliding interface, and that these differing mechanisms are directly responsible for the different frictional properties.

\section{Simulation Methods}

We employed the embedded atom method (EAM) based on a validated Ag-Cu alloy potential developed by Wu and Trinkle.\cite{wu09} The simulation results are expected to apply to Au and Au-based alloys as well because Ag and Au, both noble metals, have  similar crystal structures (i.e., both are fcc crystals with a lattice constant of 4.09~\AA~for Ag and 4.08~\AA~for Au) and mechanical properties including bulk and shear moduli, hardness, and Poisson's ratios. We have also confirmed that the simulation results of pure Au, using either the EAM potential from Foiles et al.\cite{Foiles1986} or the one from Grochola et al.,\cite{Grochola2005} are similar to those reported here for pure Ag.\cite{Argibay2017}

\begin{figure}[htb]
\centering
\includegraphics[width=0.45\textwidth]{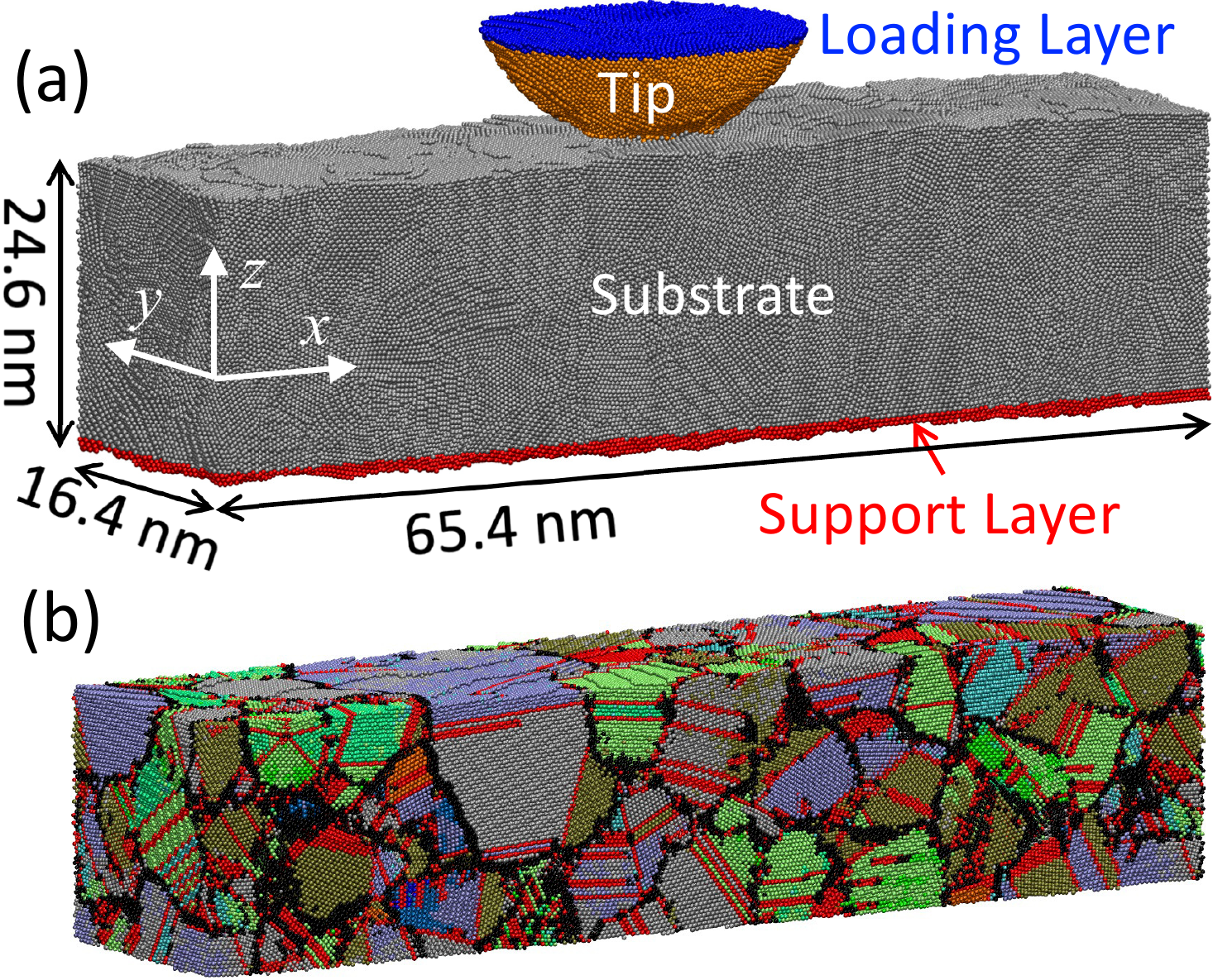}
\caption{(a) A spherical tip in contact with a rectangular cuboid substrate. (b) The nanocrystalline structure of the substrate is revealed via grain analysis.}
\label{Fig_contact_geometry}
\end{figure}

To create nanocrystalline Ag, we first prepared a bulk fcc crystal of Ag with a rectangular cuboid shape (65.4 nm $\times$ 16.4 nm $\times$ 24.6 nm). In all the MD simulations reported here, periodic boundary conditions were employed in the $x$-$y$ plane while the systems have free surfaces along the $z$-axis (see Fig.~\ref{Fig_contact_geometry} for the coordinate system). The Ag crystal was melted at 1800 K for 20 ps, and then rapidly cooled to 300 K over 100 ps to allow grains to nucleate and grow. The nucleation of grains began at the free surfaces, and then grains grew into the bulk of the slab. The cooled slab was equilibrated at 300 K for 1500 ps. To achieve a desired average grain size, the slab was heated again to 900 K over 100 ps, thermally annealed at 900 K for 500 ps, and subsequently cooled to 300 K over 100 ps. Finally, the slab was relaxed at 300 K for 2000 ps. The resulting Ag slab has an average size of crystalline grains about 5 nm, as shown in Fig.~\ref{Fig_contact_geometry}.

After the substrate was created, a tip in the shape of a spherical cap with radius of 10 nm was generated by copying a portion of the interior of the substrate. The tip created thus had an initial microstructure different from the top of the substrate. The tip was initially placed above the substrate surface. A thin layer of atoms at the bottom of the substrate and another layer at the top of the tip were used as a support layer and a loading layer, respectively, as shown in Fig.~\ref{Fig_contact_geometry}. Both layers were treated as rigid bodies during simulations. The support layer was always held stationary to hold the substrate in place. To create a series of contacts between the tip and substrate, the loading layer was displaced downward (i.e., along the $-z$ direction; see Fig.~\ref{Fig_contact_geometry} for the coordinate system) at a fixed velocity of $0.2$ m/s. To compute friction for a given contact, the loading layer was fixed in the $z$ direction and sheared along the $x$ axis at a velocity of $1$ m/s, causing the tip to slide over the substrate surface. The separation between the loading and support layers was therefore fixed during sliding. Tips possessing different microstructures and initially contacting different parts of the top surface of the substrate were also tested and yielded results almost identical to those reported here.

We conducted all MD simulations with LAMMPS using a velocity-Verlet algorithm with a time step of 1 fs to integrate the equations of motion.\cite{Plimpton1995} A layer of atoms adjacent to the support layer, which was far from the tip-substrate contact, was thermalized at 300 K with a Langevin thermostat. To further minimize its effect on friction, the thermostat was only applied to these atoms' motion in the $y$ direction, i.e. orthogonal to the direction of compression and shearing.\cite{Thompson1990} We found that the temperature at the contact rose by less than 10 K during sliding in all cases, indicating that thermally assisted grain growth was negligible. Furthermore, similar results were obtained when all the atoms were thermalized at 300 K with a Langevin thermostat in the $y$ direction. During simulations, the forces on the loading and support layers as well as those between the tip and substrate were computed to yield information on the normal load and friction.

\section{Results and Discussion}

When the tip approaches and contacts the substrate surface, the tip and substrate cold weld with an adhesive strength $\sim 4$ GPa, in reasonable agreement with experimental measurements.\cite{bhushan95,thomas93,alcantar03} With further compression, the attractive regime is followed by a repulsive force as the tip becomes more heavily deformed and embedded in the substrate. We chose a number of different separations between the tip and substrate (i.e., between the loading and support layers in Fig.~\ref{Fig_contact_geometry}) at which shear simulations were performed.

\begin{figure}[htb]
\centering
\includegraphics[width=0.45\textwidth]{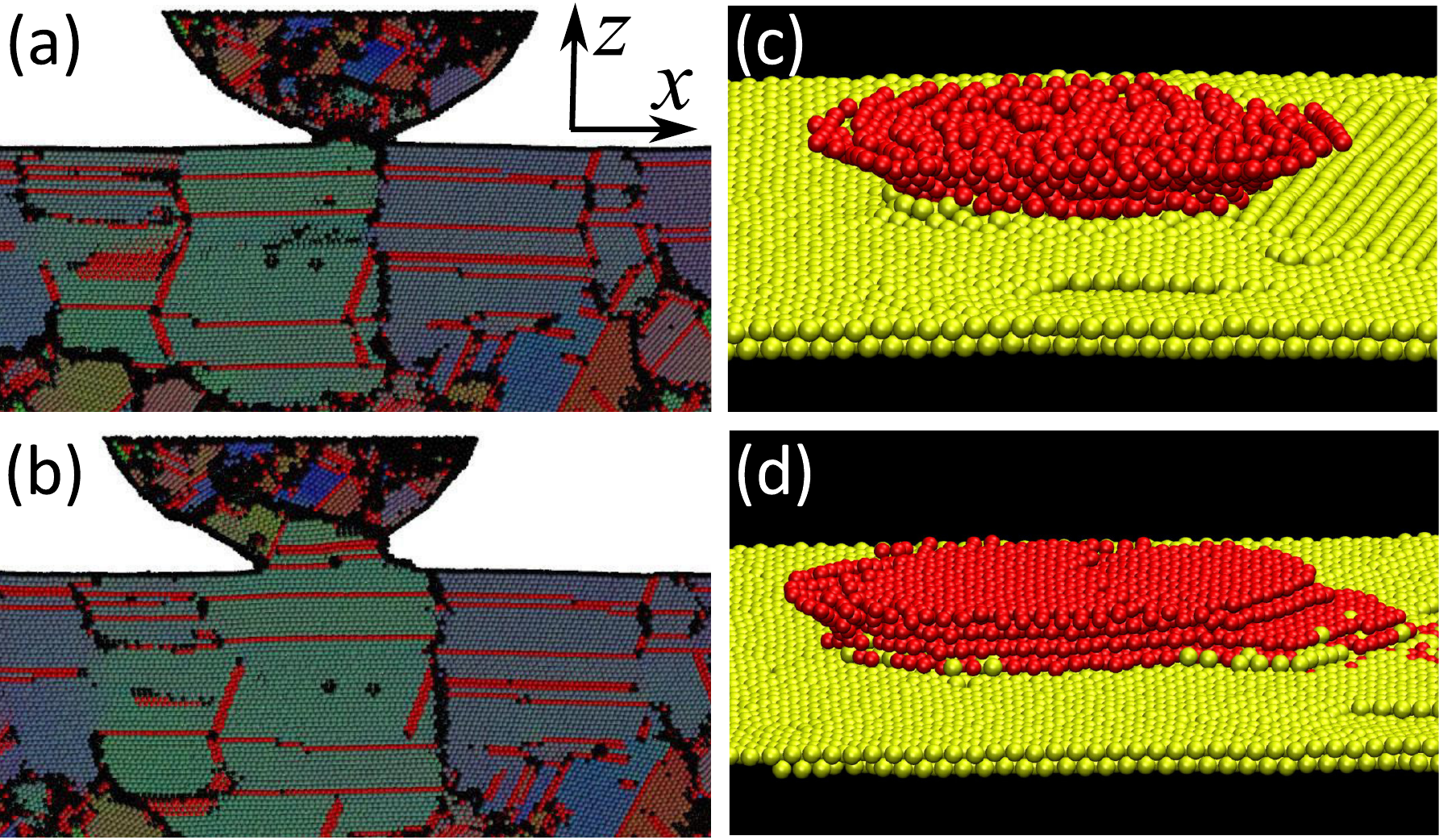}
\caption{Ag tip-slab contact (a) and (c) before sliding and (b) and (d) after 6 nm of sliding to the $-x$ direction. In (a) and (b), a thin slice in the $xz$ plane and in the middle of the contact along the $y$ direction is shown with atoms colored according to the results of grain analysis. To demonstrate the microstructural evolution, snapshots of atoms at the bottom of the tip (red) and the top of the substrate (yellow) are shown in (c) and (d).}
\label{Fig_sliding_Ag_tip}
\end{figure}

To gain insight into the structures of the nanocrystalline metals modeled here and their evolution during contact and sliding, we performed a structural analysis of atomic configurations. The local packing around each atom was analyzed with an in-house code that compared a small region around that atom to known crystalline structures (fcc, hcp, bcc). All atoms locally in an fcc environment are colored based on their grain membership. Atoms with hcp packing such as those at twin boundaries and stacking faults are colored red. All other atoms are colored black, including those at grain boundaries. The images shown in Figs. \ref{Fig_sliding_Ag_tip}(a) and (b) are obtained from such analysis. In the top and bottom regions of the substrate, twin boundaries and stacking faults are predominantly aligned parallel to the free surfaces,  indicating that $\{111\}$ planes dominate the surface texture. Before sliding, the contact between the tip and substrate is essentially a grain boundary, as shown in Fig. \ref{Fig_sliding_Ag_tip}(a). After the tip slides by 6 nm along the substrate surface, the width of the contact becomes larger and the tip's bottom undergoes microstructural reorientation to form a single grain across the interface,\cite{lu10} as shown in Fig. \ref{Fig_sliding_Ag_tip}(b). This reorientation typically occurs during the first 4 nm of sliding, a distance that is only slightly larger than the width of the initial contact. Figures \ref{Fig_sliding_Ag_tip}(c) and (d) further show the atoms at the bottom of the tip and the top of the substrate, which are colored red and yellow, respectively. The structural reorientation of atoms at the bottom of the tip and the formation of a crystalline domain in the contact zone after shear can be clearly seen.

\begin{figure}[htb]
\centering
\includegraphics[width=0.35\textwidth]{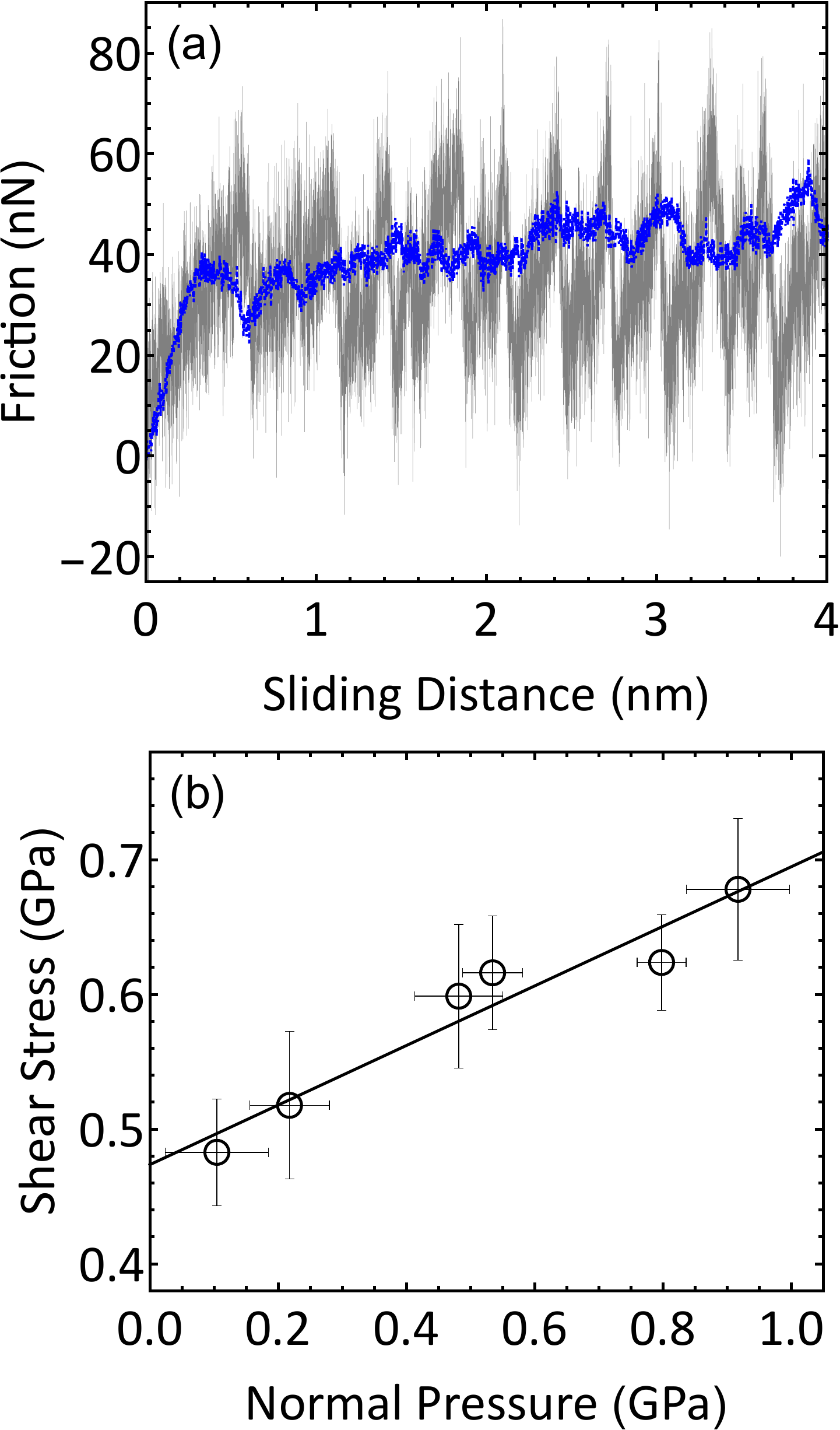}
\caption{(a) Friction force vs. sliding distance and (b) shear stress vs. normal pressure for a spherical tip sliding over a flat substrate. In (a), the gray curve shows stick-slip behavior for pure Ag while the blue curve shows no evidence of stick-slip for the Ag-Cu alloy. The data in (b) are for Ag.}
\label{Fig_fric_Ag_ela_tip}
\end{figure}

Figure \ref{Fig_sliding_Ag_tip} demonstrates that the contact has become commensurate during sliding. As a result, sliding occurs at fcc slip planes, which in this case are along the $\{111\}$ planes and in the $\langle 110 \rangle$ direction. The commensurate contact leads to a saw-tooth signal in the friction trace, with one example shown in Fig.~\ref{Fig_fric_Ag_ela_tip}(a). This demonstrates the atomic-scale stick-slip behavior that has been experimentally observed previously.\cite{gosvami11} It is by now well understood that sliding between commensurate surfaces yields high friction,\cite{he99} and in this case results in a friction coefficient $\mu=0.22$, as calculated from a linear fit to the shear stress vs. normal pressure data shown in Fig. \ref{Fig_fric_Ag_ela_tip}(b). The stress and pressure are obtained from the corresponding average friction and normal forces and the contact area computed for each contact.\cite{Cheng2010TL} The results shown in Figs. \ref{Fig_sliding_Ag_tip} and \ref{Fig_fric_Ag_ela_tip} indicate that suppressing grain growth (and thus the formation of a commensurate contact) may be a possible mechanism for maintaining low friction.\cite{he99,Ringlein2004} This understanding has motivated the work on alloys reported below.

The nanocrystalline slab of Ag shown in Fig.~\ref{Fig_contact_geometry} was used to create an alloy substrate by randomly changing 12\% of all the atoms in the slab into Cu atoms, corresponding to the composition of sterling silver.\cite{Sterling_Silver} The slab was equilibrated at 300 K for 1 ns to allow the crystalline domains and boundaries to relax. After equilibration, an Ag-Cu alloy tip in the shape of a spherical cap was copied from the interior of the Ag-Cu slab and a tip-substrate contact similar to the one in Fig.~\ref{Fig_contact_geometry} was created. A protocol similar to that discussed earlier for the pure Ag system was used to create a series of contacts with the Ag-Cu alloy at different separations between the loading and support layers that were subsequently used in sliding simulations to compute friction.

\begin{figure}[htb]
\centering
\includegraphics[width=0.45\textwidth]{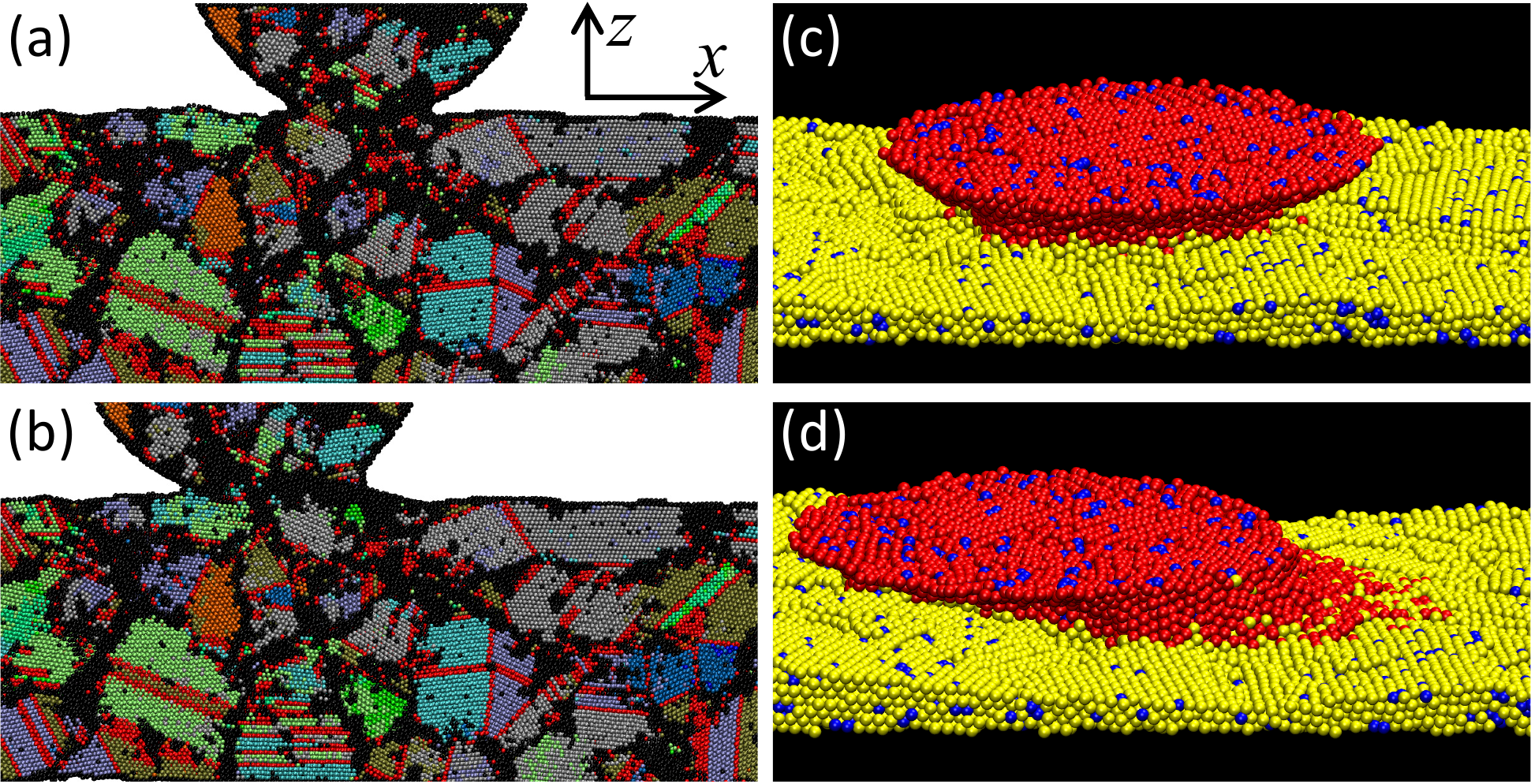}
\caption{Ag-Cu tip-slab contact (a) and (c) before sliding and (b) and (d) after 6 nm of sliding to the $-x$ direction. In (a) and (b), a thin slice in the $xz$ plane and in the middle of the contact along the $y$ direction is shown with atoms colored according to the results of grain analysis. To demonstrate the lack of microstructural evolution, snapshots of atoms at the bottom of the tip (red for Ag and blue for Cu) and the top of the substrate (yellow for Ag and blue for Cu) are shown in (c) and (d).}
\label{Fig_sliding_AgCu_tip}
\end{figure}

An example of the tip-substrate contact for the Ag-Cu alloy is shown in Fig.~\ref{Fig_sliding_AgCu_tip}(a), with the image showing a thin slice of the contact with atoms colored based on grain analysis. The image shows that the average grain size is smaller and the grain boundaries are wider for the Ag-Cu alloy. This is because after changing 12\% of the atoms to Cu, the nanocrystalline substrate did not undergo any further heat treatments except for an reequilibration process at 300 K. Some grains were therefore effectively fragmented by the Cu atoms. These atoms also tended to diffuse toward and accumulate at grain boundaries, increasing their width. It is also clear that before sliding the contact interface is essentially a grain boundary. The same contact after sliding the tip with respect to the substrate by 6 nm is shown in Fig.~\ref{Fig_sliding_AgCu_tip}(b). In the case of the alloy contact, the width of the contact becomes larger during sliding, similar to the pure metal case, but grain growth is not observed in the contact region. Instead, the contact zone remains a grain boundary after shear. Snapshots of the atoms in the contact region, shown in Fig.~\ref{Fig_sliding_AgCu_tip}(c) before sliding and in Fig.~\ref{Fig_sliding_AgCu_tip}(d) after sliding, confirm that although the bottom of the tip is severely deformed during sliding, no crystalline domains are formed at the contact. Note that the surface of the substrate shows an extremely refined grain structure.

We have computed the shear stress of the Ag-Cu alloy tip sliding over the Ag-Cu substrate at a series of separations between the loading layer at the top of the tip and the support layer at the bottom of the substrate. As the separation is reduced, the average normal pressure and shear stress become larger, but the calculated work of adhesion between the alloy tip and substrate is almost twice that of pure Ag. This high adhesion in the Ag-Cu alloy makes the tip prone to deformation during sliding, as shown in Fig.~\ref{Fig_sliding_AgCu_tip}(d). The consequence is that, during sliding under a fixed separation between the loading and support layers, the normal pressure between the tip and substrate starts to decrease after initial sliding, and a friction vs. normal load curve cannot be obtained. However, it is clear that the friction trace does not exhibit any stick-slip behavior like that for Ag. An example of friction force vs. sliding distance for the Ag-Cu tip-substrate contact is shown in Fig.~\ref{Fig_fric_Ag_ela_tip}(a), together with an example for Ag. Clearly there is no stick-slip signal for the alloy contact, confirming the disordered nature of the interface between the alloy tip and substrate during sliding.

\begin{figure}[htb]
\centering
\includegraphics[width=0.3\textwidth]{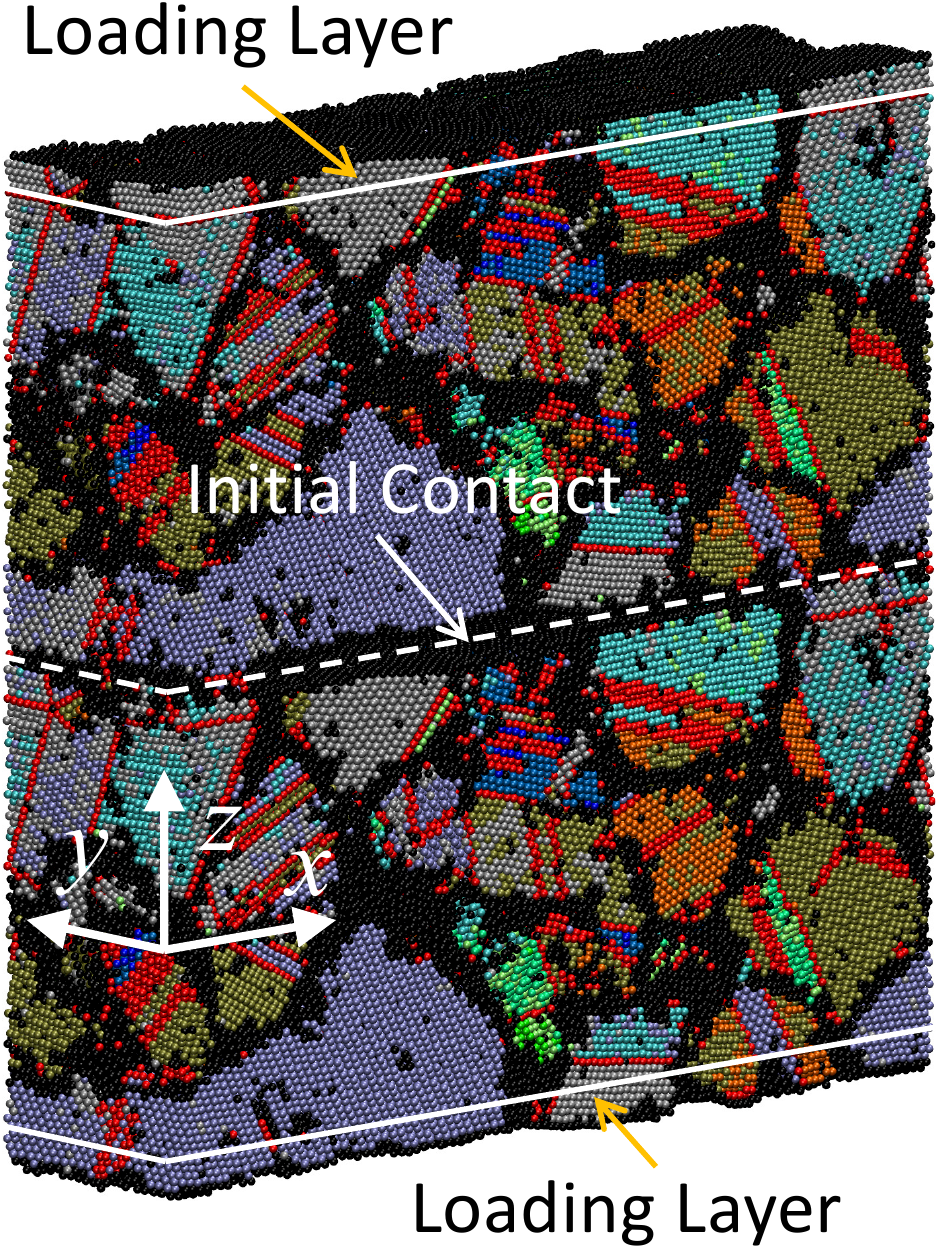}
\caption{A contact between two Ag-Cu slabs.}
\label{Fig_FOF_contact}
\end{figure}

To have a direct comparison between pure metals and alloys, we resort to a contact between two slabs with each slab made of either pure Ag or the Ag-Cu alloy. A similar geometry was also used by Romero et al. to investigate the evolution of plastic deformation in nanocrystalline metals sliding against each other.\cite{Romero2014} Here the size of each slab is 32.7 nm $\times$ 8.2 nm $\times$ 16.4 nm. One such contact for the Ag-Cu alloy is shown in Fig.~\ref{Fig_FOF_contact}. A thin layer of atoms at the top of the top slab and another layer at the bottom of the bottom slab were designated as loading layers. In the initial configuration, the two slabs were separated by a small gap. To bring the slabs into contact, the two loading layers were moved toward each other along the $z$ axis at a fixed velocity of $0.1$ m/s. The resulting force vs. separation curves (not shown) show that both pure Ag and the Ag-Cu alloy exhibit very similar work of adhesion for these slab-on-slab contacts. To compute friction, the two loading layers were held at a fixed separation and sheared in opposite directions along the $x$ axis at a velocity of $1$ m/s, causing the two slabs to slide against each other. The sliding plane always began close to the initial contact between the two slabs (Fig.~\ref{Fig_FOF_contact}) but the evolution of its location over time was different for pure Ag and the Ag-Cu alloy. The normal and lateral force at the interface between the slabs were computed as sliding evolved to yield the normal pressure and shear stress.

 \begin{figure}[htb]
\centering
\includegraphics[width=0.4\textwidth]{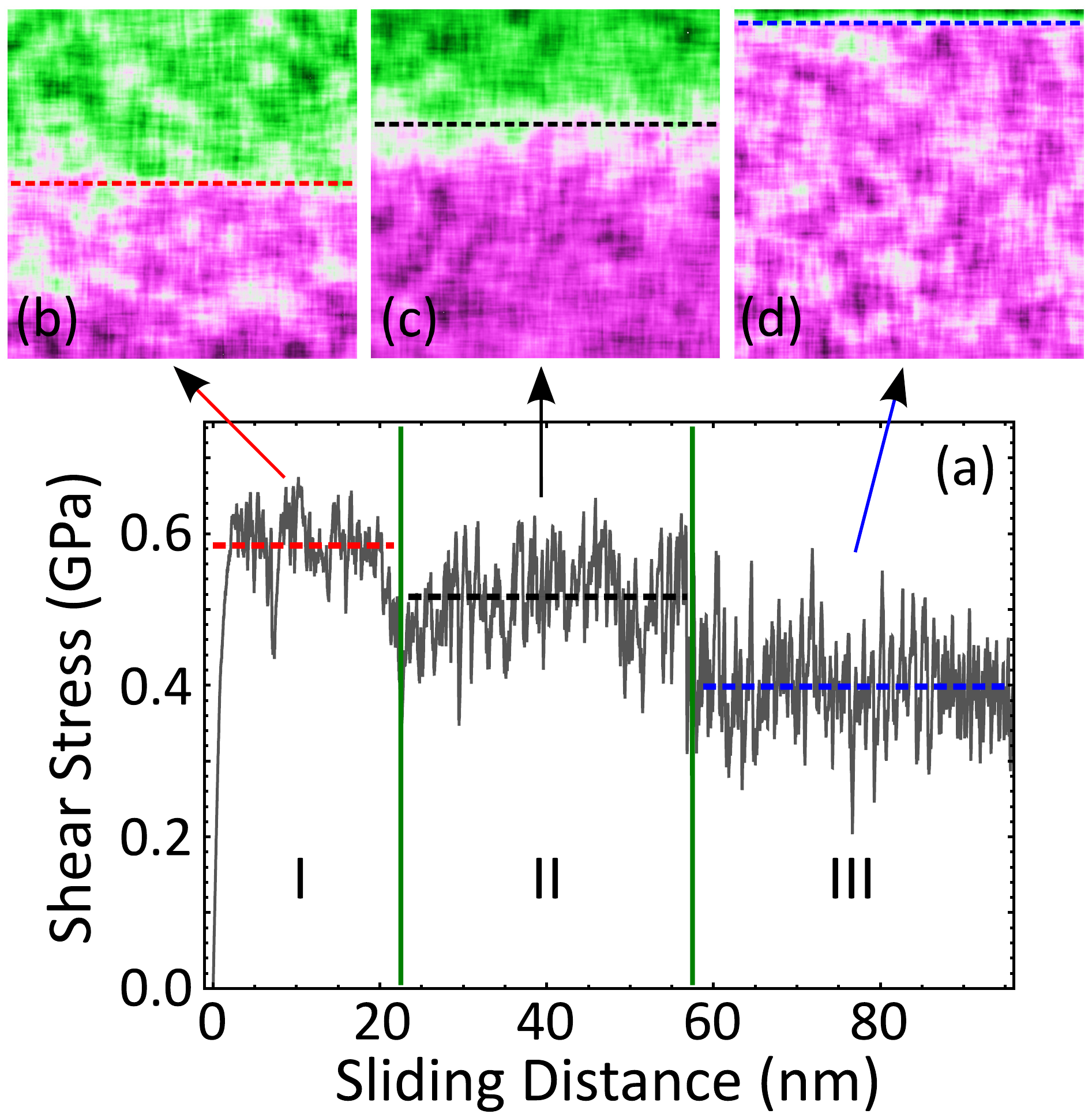}
\caption{Contact between two Ag slabs under shear:
(a) Friction trace as a function of sliding distance showing three distinct regimes: I and (b), high friction when sliding occurs primarily along slip planes near the initial interface between the two slabs; II and (c), intermediate friction as the sliding plane moves toward the upper loading layer; III and (d), low friction when sliding occurs primarily at the boundary of the loading layer. In (b)-(d), the local shear velocity is shown with green-pink density plots with green indicating $1$ m/s and pink indicating $-1$ m/s, and the dashed line in each plot indicates the location of the slip plane.}
\label{Fig_sliding_Ag_FOF}
\end{figure}

Figure \ref{Fig_sliding_Ag_FOF}(a) shows the calculated shear stress as a function of sliding distance for the pure Ag system, with three distinct regimes. The regimes arise because the nanocrystalline Ag slabs undergo significant grain coalescence and growth under shear \cite{Argibay2017} and the sliding plane moves toward one of the rigid loading layers. The movement of the slip plane during sliding is clearly demonstrated in the plots of local shear velocity shown in Figs.~\ref{Fig_sliding_Ag_FOF}(b)-(d). A detailed analysis indicates that initially, slip primarily occurs through grain boundaries at the middle of the system, close to the initial interface between the two slabs. As sliding continues, grains rotate, merge, and grow, causing the shear stress to increase steadily. In regime I sliding is primarily accommodated through twin boundaries and stacking faults and friction remains high. However in regime II the sliding plane gradually moves toward one of the fixed surfaces because of grain growth across the initial contacting interface (similar to Fig. \ref{Fig_sliding_Ag_tip}(b)) and the vertical propagation of twin boundaries and stacking faults. Similar localization of the sliding shear and movement of the slip plane were also observed in the MD simulations of Romero et al. for pure Fe and Cu.\cite{Romero2014} In regime III sliding primarily occurs at the interface between the polycrystalline Ag and the fixed loading layer used to impose shear. Because grain growth cannot progress into the fixed layer, this interface is essentially a sequence of grain boundaries traversing the system, resulting in a transfer film that grows between the loading layer and the bulk polycrystal. The transitions in the shear stress, shown in Fig. \ref{Fig_sliding_Ag_FOF}(a), are a direct consequence of the movement of the sliding plane.

 \begin{figure}[htb]
\centering
\includegraphics[width=0.4\textwidth]{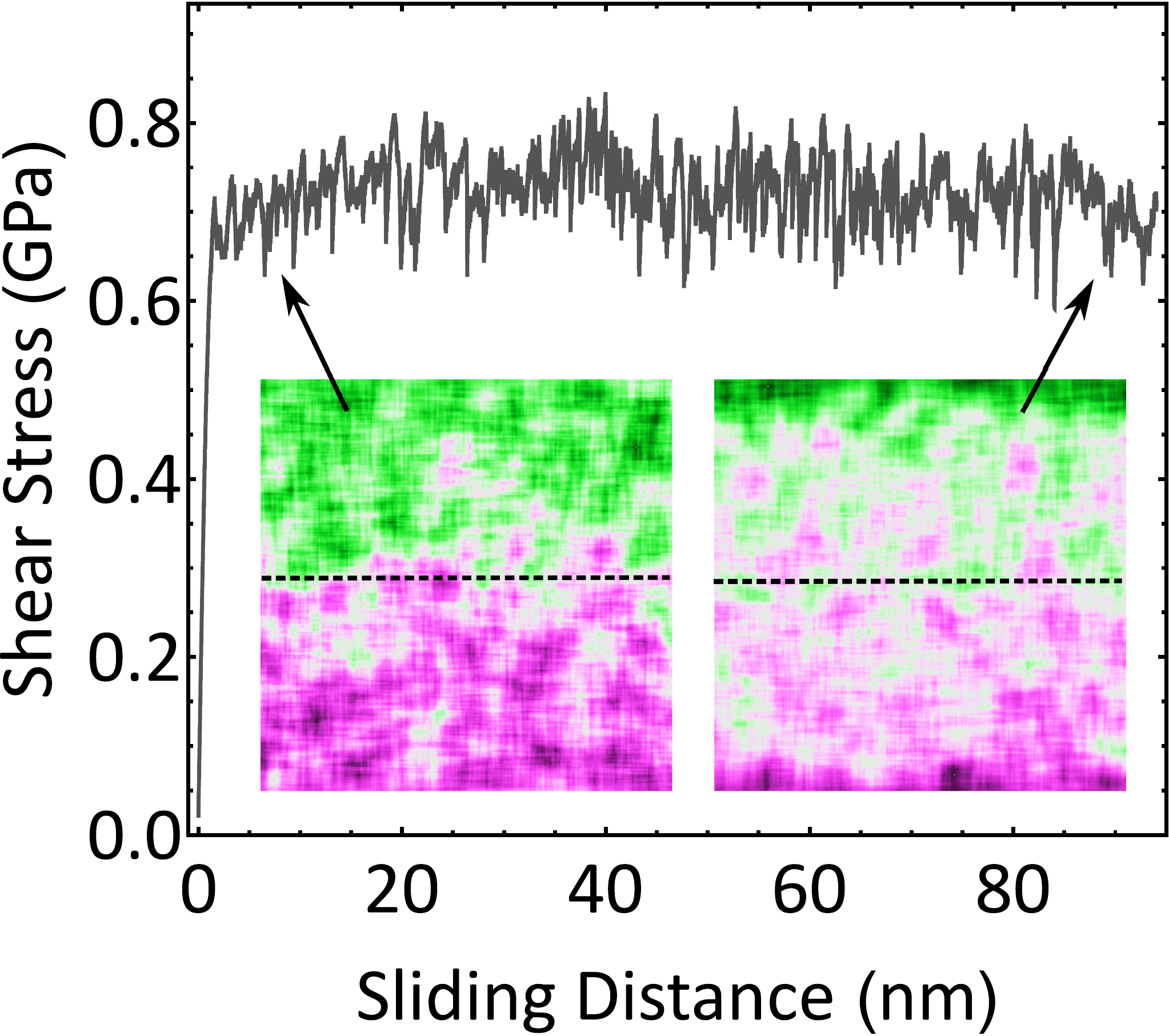}
\caption{Friction trace as a function of sliding distance for a contact between two Ag-Cu slabs under shear. Inset: green-pink density plots (with the same color scale as in Fig.~\ref{Fig_sliding_Ag_FOF}) of local shear velocity showing the location of the slipping plane (dashed lines) in the early and final stage of the sliding process.}
\label{Fig_sliding_AgCu_FOF}
\end{figure}

These results are in stark contrast to the progression of shear in the Ag-Cu contact. As shown in Fig.~\ref{Fig_sliding_AgCu_FOF}, the shear stress as a function of sliding distance does not show any transitions. The insets of Fig.~\ref{Fig_sliding_AgCu_FOF} show the density plots of local shear velocity, indicating that the slip plane remains near the initial contact between the two slabs, and grain analysis reveals that sliding is accommodated by grain boundaries in the system. In some cases, the average location of the slip plane moves away from the initial interface as its location can fluctuate over time, but sliding always occurs at grain boundaries that traverse the system in the direction of sliding. In this case, the presence of Cu atoms prevents the grains in the Ag-Cu alloy from growing \cite{Argibay2017} by stabilizing grain boundaries and allowing the slabs to slide over each other. As sliding in the Ag-Cu alloy occurs at a disordered interface, a smaller friction coefficient is expected. Indeed, for the slab-on-slab contacts, the friction coefficient is about $0.076$ for pure Ag (based on the friction trace in regime I in Fig.~\ref{Fig_sliding_Ag_FOF}(a)) and $0.027$ for the Ag-Cu alloy.\cite{Argibay2017} At the atomic scale, the change in the friction efficient is a result of the different sliding mechanisms accessible to pure Ag and Ag-Cu alloys. In the former, grain growth under shear leads to the formation of a commensurate contact and high friction while in the latter, grain growth is suppressed and shear is accommodated by disordered grain boundaries, resulting in lower friction.

\section{Conclusions}

The simulation results reported here demonstrate that shear along grain boundaries, where the structure is nearly amorphous, results in low friction. The presence of incompatible elements in an alloy can prevent grain growth through stabilization of the grain boundaries that accommodate shear and plastic deformation. In pure metals, grain growth can be significant under stress and as a result, the slip planes are commensurate contacts with high friction. These mechanisms have also been proposed for previous experiments on Ag and Cu \cite{Argibay2017} as well as nanocrystalline Ni \cite{Prasad2011} and single crystal Ni with a sliding-induced ultra-nanocrystalline surface layer.\cite{Prasad2020}

The implication of our finding is that variations of materials properties, such as ductility or hardness, are not responsible for the different friction coefficients in metals and alloys. Rather, alloying (and the formation of composites) breaks the ability of the material to form a commensurate contact, seen in pure metals as the formation of a single, oriented grain across the cold-welded interface. The potential to form low-friction alloys with less consideration paid to other material properties could lead to better performing metallic contacts with high conductivity, high wear resistance, and low friction.\cite{Curry2018} Surface engineering to prevent a $\{111\}$ dominated surface texture is also an option for preventing dislocation-mediated plasticity, thereby enforcing grain boundary sliding and lowering friction.\cite{Prasad2020}

The change in hardness upon alloying in experimental systems is generally ascribed to the Hall-Petch effect, in which decreasing grain size leads to increased hardness. Alloying stabilizes grain boundaries through Zener pinning or solute drag, preventing grain growth. The correlation between hardness and grain size is characterized by two regimes, with plasticity in grains larger than $\sim 10$ nm primarily occurring through dislocation activity, while in smaller grains plasticity is dominated by grain boundary sliding.\cite{lo79,Swygenhoven1997,Swygenhoven2002Science,Swygenhoven2006,Schiotz1998,Schiotz2003,Schiotz2004,Shan2005,Wolf2005,Ivanisenko2009,Cordero2016,Guo2018PRL,Gupta2020} Essentially the same effect is seen in our simulations, with only minimal differences. While experimentally the solute often segregates to grain boundaries, the solute atoms in the Ag-Cu alloy modeled here are well dispersed throughout the metal matrix. Regardless, alloying still prevents grain growth in both situations. In our simulations, grains are initially small for both the pure metal and the alloy, with significant grain growth seen in the pure metal only. Experimentally it has been shown that pure Au systems have significantly larger grains compared to composites,\cite{argibay13} and that low friction in metals is often associated with the formation of an ultra-nanocrystalline region at the sliding interface after run-in \cite{Prasad2011}. Rather than the increased hardness leading to lower friction in the alloys, it is more accurate to state that the same mechanism leading to the increase in hardness is responsible for the lowering of friction; namely, changes in both hardness and friction are controlled by grain size and the dominance of either dislocation-mediated plasticity or grain boundary sliding.

\section*{Acknowledgements}
S.C. acknowledges Advanced Research Computing at Virginia Tech (URL: http://www.arc.vt.edu) for providing computational resources and technical support that have contributed to the results reported within this paper. S.C. also gratefully acknowledges the support of NVIDIA Corporation with the donation of the Tesla K40 GPUs used for this research. This work was performed, in part, at the Center for Integrated Nanotechnologies, an Office of Science User Facility operated for the U.S. Department of Energy Office of Science. This work was funded by the Laboratory Directed Research and Development Program at Sandia National Laboratories, a multimission laboratory managed and operated by National Technology and Engineering Solutions of Sandia, LLC., a wholly owned subsidiary of Honeywell International, Inc., for the U.S. Department of Energy's National Nuclear Security Administration under contract DE-NA0003525. This paper describes objective technical results and analysis. Any subjective views or opinions that might be expressed in the paper do not necessarily represent the views of the U.S. Department of Energy or the United States Government.


\end{document}